\documentclass{article}

\usepackage{PRIMEarxiv}

\usepackage[utf8]{inputenc} 
\usepackage[T1]{fontenc}    
\usepackage{hyperref}       
\usepackage{url}            
\usepackage{booktabs}       
\usepackage{amsfonts}       
\usepackage{nicefrac}       
\usepackage{microtype}      
\usepackage{lipsum}
\usepackage{fancyhdr}       
\usepackage{graphicx}       
\graphicspath{{media/}}     
\usepackage{amsmath}
\usepackage{float}

\title{Unveiling Activity-Travel Patterns through Topological Data Analysis
}

\author{
  Nuoxian Huang \\
  Behavioral and Spatial AI Lab \\
  Peking University Shenzhen Graduate School \\
  \texttt{huangnx0927@stu.pku.edu.cn} \\
   \And
  Yulin Wu \\
  Behavioral and Spatial AI Lab \\
  Peking University Shenzhen Graduate School \\
  \texttt{cover.wu@stu.pku.edu.cn} \\
}

\begin{document}
\maketitle

\begin{abstract}
In the context of rapid urbanization, understanding the patterns of urban residents' activities and mobility is crucial for optimizing transportation systems and enhancing urban management efficiency. This study addresses the limitations of traditional travel analysis methods in handling high-dimensional and large-scale spatiotemporal data by incorporating Topological Data Analysis (TDA) techniques, specifically using persistent homology. This method allows for the extraction of information from the topological structure of data, enabling the effective identification and analysis of complex spatiotemporal behavior patterns without reducing the data's dimensionality. We utilized mobile signaling data from a community in Shenzhen, which includes detailed geographic and temporal information, providing an ideal sample for analyzing urban residents' behavior patterns. Using our pattern mining framework, we successfully identified five main patterns of residents' activities and travel, revealing daily behavioral habits and reflecting the activity heterogeneity among residents with different socio-economic attributes. These findings not only assist urban planners in better session design but also provide new characteristics for predictive mobility models.
\end{abstract}


\section{Introduction}
The study of urban transportation and travel behavior has been extensively researched over the past decades \cite{golob2003structural,gotschi2017towards}. These studies not only help us understand the daily travel patterns of urban residents but also provide important data and theoretical support for urban planning and traffic management. Traditional trip-based models primarily focus on the origin, destination, and route choice of travel, failing to fully consider the complexity of travel behavior as a derivative demand of human activities. Consequently, there has been an increasing focus on activity-based approaches \cite{kitamura1988evaluation}, which suggest that human travel behavior should be understood from the perspective of activities. By simulating individuals' daily activities and travel decision processes, these approaches offer more realistic behavior reproduction and policy evaluation \cite{dong2006moving,liu2021similar,shiftan2008use}. This research paradigm has prompted a shift in transportation planning from viewing residents as a whole towards focusing on the diverse behaviors of different groups, as different activity needs trigger different travel choices. However, it remains challenging for transportation planning models to obtain or predict the travel behaviors of all individuals within a simulated area. Thus, exploring urban residents' patterns of activity and travel, and obtaining representative homogeneous patterns becomes particularly important in transportation planning.

In the past, due to the limit of mobile communications, the internet, and wireless positioning technologies, acquiring data on urban residents' travel behavior was extremely challenging, often gathered through observations, travel and activity diaries survey. These methods were costly, covered small samples, and spanned short periods, making it difficult to observe and record residents' spatial movements on a large scale and over extended duration. In recent years, with technological advancements, it has become possible to obtain high spatial-temporal precision, massive, long-term series of spatiotemporal behavioral trajectory data. The improvement in data quality and availability, developments in new technologies, and interdisciplinary integration have provided a solid foundation for the study of residents' travel behavior, facilitating the quantitative analysis of travel behavior \cite{huang2018tracking,reades2016finding}. In this context, the continuous mining and tracking of spatiotemporal behavioral trajectory data to uncover underlying pattern features has become a research focus. The group behavior patterns help to understand individual social connections and their preferences for the environment \cite{zhang2017detecting}.

Time geography initially proposed a systematic analytical framework based on individual spatiotemporal behavior trajectories, primarily utilizing visualization methods such as space-time path and space-time prism to explore behavior patterns in personal activity diaries data. While visual expressions based on time geography are beneficial for uncovering hidden patterns and their spatiotemporal interactions within trajectory data, they struggle with computation in large-sample datasets and direct integration into transportation planning efforts. To identify regularities in residents' activity patterns within time geographic expressions and provide new insights, we have developed a research framework that combines topological data analysis with sequence analysis for studying activity-travel patterns. By clustering based on individuals' activity-travel characteristics, our goal is to identify diverse activity-travel patterns of urban residents, thereby designing more just services and policies to ensure the needs of different groups are met. Additionally, the pattern we capture can serve as encoding or representations of human mobility, offering predictive features for a range of human mobility prediction models in deep learning.

The structure of the remainder of this paper is as follows. In Section 2, we provide a literature review on activity-travel pattern mining and topological data analysis. Section 3 describes the data used in this study and the analytical method and overall framework. In Section 4, we present the clustering of daily activity patterns and the discovery of their associated sociodemographic characteristics. Finally, in Section 5, we discuss and summarize our research findings.

\section{Literature review}
\subsection{Activity-travel pattern}
The types of patterns contained in spatiotemporal behavioral trajectory data are diverse and can generally be categorized into periodic patterns \cite{yang2001infominer}, frequent patterns\cite{yin2015mining}, gathering patterns\cite{kalnis2005discovering}, semantic patterns \cite{giannotti2007trajectory}, and activity-travel patterns\cite{zhang2017detecting}. These types are not entirely independent and can often be mined together; this study primarily focuses on periodic (temporal) activity-travel patterns.

Periodic patterns refer to subsequences that repeatedly occur in spatiotemporal behavioral trajectories, often used to describe behaviors with temporal regularities. The identification of these patterns suggests that moving objects tend to follow the same or similar mobility patterns within specified time intervals. This approach primarily focuses on the intensity of time-related human mobility, rather than spatial density. Research on periodic patterns can be categorized into two methods: (1) Given a period, examining the characteristics of mobility within this fixed period. For instance, Mamoulis et al.\cite{mamoulis2004mining} analyzed temporally regular movements by transforming trajectories within each cycle into region-based sequences and mining the periodic patterns therein. (2) Mining the periodic characteristics of moving objects' spatiotemporal trajectories without pre-setting the length of the time period. For example, Li et al.\cite{li2016probabilistic} introduced a probabilistic model-based hybrid periodic pattern mining method to address multiple mixed periodic behaviors; Yuan et al.\cite{yuan2017pred} employed a Bayesian non-parametric model to uncover periodic mobility patterns by jointly modeling geographic and temporal information. Moreover, there are approaches that integrate multi-dimensional information of trajectories for mining periodic patterns, as seen in the work of Bermingham and Lee\cite{bermingham2020mining} and Zhang et al. \cite{zhang2018hierarchical,zhang2019mining}. However, these methods generally focus on mining periodic patterns based on individual movement behaviors, whereas Shi et al.\cite{shi2019mining} proposed the GPMine algorithm for mining group-based periodic movement patterns.

The majority of existing studies on mobility pattern mining focus on identifying groups with similar patterns and then examining the socio-economic attributes associated with these patterns, which aligns with the goals of this research. However, current methods exhibit certain limitations in processing spatiotemporal trajectory features. Firstly, the understanding of spatiotemporal trajectories appears dichotomous: existing methods either separate the multidimensional attributes of trajectories or overly emphasize these features, neglecting the holistic characteristics. Secondly, the dynamic of travel behavior is not adequately considered, leading to a lack of analysis on long-term continuous spatiotemporal data. To address these issues, this study proposes a novel pattern mining framework that integrates topological and geometric methods. Using continuous mobile signaling data in one month from a community in Shenzhen, this study employs a framework that incorporates Topological Data Analysis (TDA). TDA is advantageous for its ability to preserve dimensionality and its robustness against noise \cite{chazal2021introduction}. It provides a new approach for representing and analyzing the rhythmic features and behavioral patterns found in long-term, multidimensional trajectory data. 

\subsection{Topological Data Analysis}
Topological Data Analysis (TDA) employs techniques from topology to analyze complex and high-dimensional data sets, extracting insights through the inherent shapes and structures of the data. There are three key topological concepts that enable pattern extraction through shape: being coordinate-free, invariant under "small" deformations, and providing compressed representations of shapes \cite{lum2013extracting}. Within TDA, the mapper algorithm and persistent homology are two prevalent key techniques that contribute unique insights to data analysis.

Persistent homology primarily aims to analyze the stability of topological features within a dataset as scales change \cite{edelsbrunner2008persistent}. In practice, this often involves constructing a multi-scale topological object—such as a simplicial complex—that progressively "grows" based on the data. As the scale changes, certain topological features appear and disappear; persistent homology tracks these changes and analyzes the lifespan of these features. This method can be applied to data in the form of point clouds, networks \cite{feng2020spatial}, and time series \cite{ravishanker2021introduction}, and is becoming an efficient solution for data analysis across various disciplines. Specifically, in urban traffic planning, the application of persistent homology is currently focused on the study of traffic networks. For instance, Wu et al.\cite{wu2017congestion} used persistent barcode, a tool in persistent homology, to visualize the connectivity of traffic networks and illustrate the robustness against congestion. Additionally, Feng and Porter\cite{feng2020spatial} employed persistent homology for urban analysis by characterizing city morphology based on street networks. This approach provides a powerful tool for understanding complex spatial structures and their evolution over time, offering insights into urban dynamics and planning.

This study primarily utilizes persistent homology to analyze activity travel patterns, yielding several advantages: (1) It offers a method to succinctly capture the full spectrum of data without the need for dimensionality reduction, thereby addressing the limitations associated with information loss due to dimension reduction. (2) Persistent homology can identify features that standard techniques cannot detect. (3) The technique has stability against noise, enabling it to distinguish between short-lived persistent features and transient noise that only appear on a small scale. This capability enhances the precision and reliability of analyzing complex data sets.

\section{Data and methodology}
\subsection{Data}
\subsubsection{Research region and data overview}
This study is based on mobile signaling data provided by a major telecommunications operator in China, including mobile events such as calls and text messages. The data includes user IDs and the physical locations of cellular base stations. Given the massive volume of city-wide data, and this study focuses more on innovative analytical methods, only data from a specific community in Henggang Street, Longgang District, Shenzhen, in August 2019 was selected as the samples. Individual activity trajectories were classified into two types: stays and trips. A stay refers to an individual remaining within a certain distance threshold for a specified duration, whereas a trip denotes movement between stay points\cite{yang2021detecting}.

\begin{figure}
  \centering
  \includegraphics[width=0.4\textwidth]{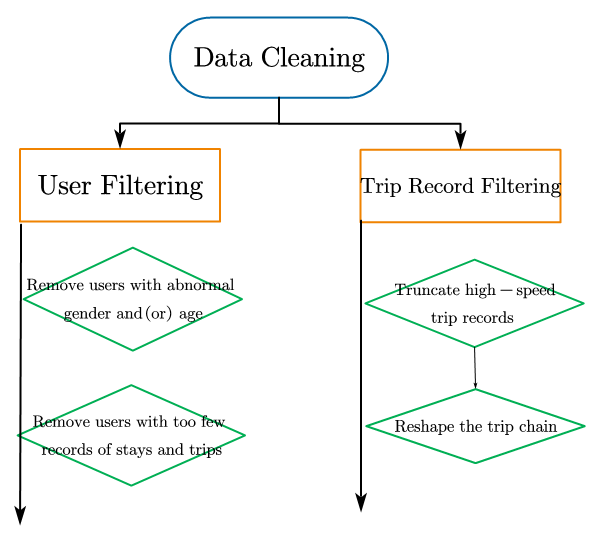}
  \caption{Data cleaning workflow}
  \label{fig:data_cleaning}
\end{figure}

The detailed content of the data used is as shown in Table~\ref{tab:data_description}. Moreover, the proactive, real-time collected mobile signaling data may be affected by noise, leading to occasional missing user attributes, unstable residential locations, and abnormal travel speeds and stay locations. Therefore, the study initially undertook a thorough data cleaning process, primarily involving user filtering and travel record selection, as illustrated in Figure~\ref{fig:data_cleaning}.

\begin{table} [ht]
 \caption{Data Structure Description}
 \label{tab:data_description} 
  \centering
  \begin{tabular}{lll}
    \toprule
    \textbf{Column} & \textbf{Data type} & \textbf{Description} \\
    \midrule
    pid      & string    & Unique user ID, identification only \\
    \midrule
    \multicolumn{3}{l}{\textbf{Activity and Travel Data}} \\
    \midrule
    date     & datetime  & Date of activity \\
    t\_start  & datetime  & Activity start time \\
    t\_end    & datetime  & Activity end time \\
    longitude & float     & Longitude of activity location \\
    latitude  & float     & Latitude of activity location \\
    ptype     & category  & Type of activity (0-home; 1-travel; 2-work or study; 3-other) \\
    \midrule
    \multicolumn{3}{l}{\textbf{Personal Attributes Data}} \\
    \midrule
    age       & category  & User age group, e.g., '5' for ages 19-24 \\
    gender    & category  & User gender (01-male; 02-female; 03-unknown) \\
    arpu      & float     & Monthly communication cost \\
    brand     & string    & Mobile phone model \\
    \bottomrule
  \end{tabular}
  \label{tab:personal_data}
\end{table}

\subsubsection{Exploratory data analysis}
\begin{figure} [H]
  \centering
  \includegraphics[width=0.8\textwidth]{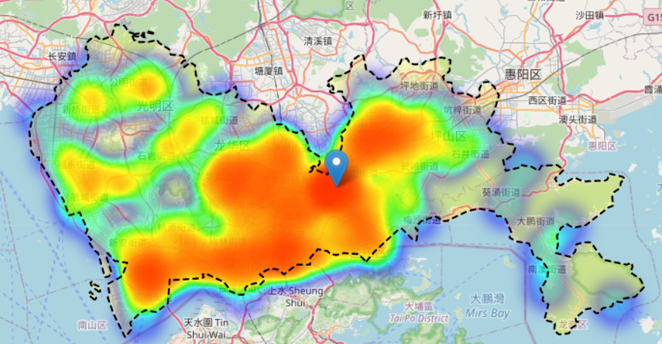}
  \caption{Community residents' activity distribution}
  \label{fig:acvtivity_heatmap}
\end{figure}

Based on the frequency of stays, Figure~\ref{fig:acvtivity_heatmap} illustrates the heatmap of activity distribution of the residential living in the surveyed community. Blue location markers indicate the community's position, serving as the main anchor point of activities and travel. The heatmap shows the highest values within approximately a 5-kilometer radius centered on this main anchor point, indicating the most frequent stays. Additionally, beyond the main anchor in Longgang District, the residents' activity areas are primarily distributed in Nanshan, Futian, and Luohu Districts. In contrast, developing areas like Guangming and Dapeng New Districts see sparse activity.

\begin{figure}[H]
  \centering
  \begin{minipage}{0.5\textwidth}
    \centering
    \includegraphics[width=\textwidth]{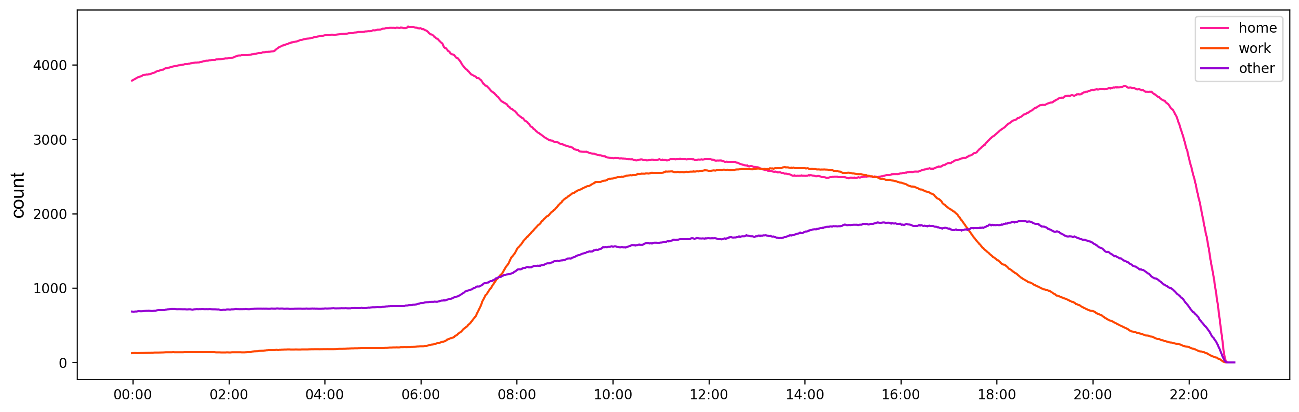}
    \caption{Daily statistics of residents' activity types}
    \label{fig:daily_activity_types}
  \end{minipage}%
  \begin{minipage}{0.5\textwidth}
    \centering
    \includegraphics[width=\textwidth]{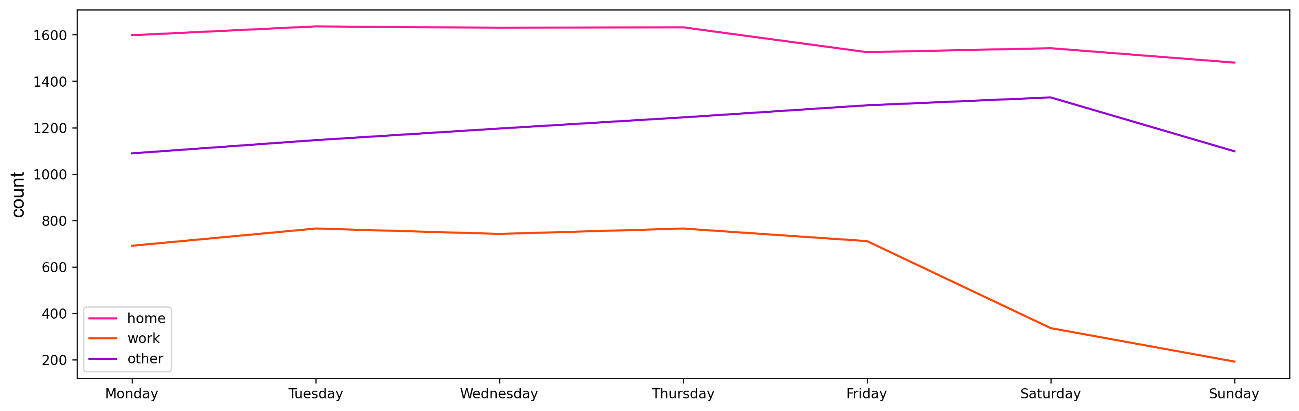}
    \caption{Weekly statistics of residents' activity types}
    \label{fig:weekly_activity_types}
  \end{minipage}
\end{figure}

The dataset also includes information on the time utilization of mobile users, allowing us to observe details such as the sequence, duration, and timing of activities, which are of particular interest to us. Figure~\ref{fig:daily_activity_types} presents a daily statistical analysis of activity frequency by minute, based on the dataset. As depicted, the "home" activity type dominates in terms of sample count, occurring mostly between 00:00 to 07:00 and 19:00 to 24:00. There is a noticeable decline from 07:00 to 10:00, with the lowest frequency occurring from 10:00 to 18:00 during the day. Conversely, the "work" activity type peaks between 10:00 and 18:00, aligning closely with the typical daily commuting rhythms of residents. This pattern confirms the validity of the collected data.

Figure~\ref{fig:weekly_activity_types} illustrates the frequency of various types of activities occurring each day of the week within the dataset. It is evident that "home" remains the most common activity type, followed by "other" activities, with no significant variation between them throughout the week. The "work" activity type shows the least frequency, likely because some mobile users do not need to work, whereas almost everyone engages in "home" and "other" activities. Moreover, the occurrence of "work" activities displays a clear daily variation, with significantly higher frequencies on weekdays compared to weekends.

\subsection{Methodology}
This study focuses on mining patterns of activity and travel based on a categorized time series of activity and travel types, similar to the research of \cite{millward2019activity} and \cite{song2021visualizing}. Formally, let \( X = \{x_{n,t}\} \), where \( t = 1, \dots, T \) and \( n = 1, \dots, N \) represent the set of \( N \) categorized time series, each of length \( T \). Each data point within these series is represented as one of \( J \) categories. Specifically, through processing and transformation of data, this study generates long-term activity and travel sequences on a monthly basis, with a data granularity of one minute. Thus, the dataset studied has \( N = 192 \), \( T = 40,320 \), and each data point \( x_{n,t} \) is represented as
\[
x_{n,t} = 
\begin{cases} 
0, & \text{visit (other activities)} \\
1, & \text{at home} \\
2, & \text{work} \\
3, & \text{trip}
\end{cases}
\]
\subsubsection{Topological analysis of activity-travel sequences}
In the first step of the topological analysis, we employ the Fast Walsh-Fourier Transform (FWT) to convert the time series \(X\) into their frequency domain representations, which proves beneficial for capturing sequency patterns in categorized time series \cite{chen2019clustering}. Although the FWT analysis captures the sequential characteristics of the time series and effectively eliminates redundancy, it is challenging to use directly as a feature for time series clustering. Therefore, in the second step, we utilize persistent homology techniques from TDA to analyze the spectrum and generate persistence features using persistent landscapes.

\begin{figure}[H]
  \centering
  \includegraphics[width=0.3\textwidth]{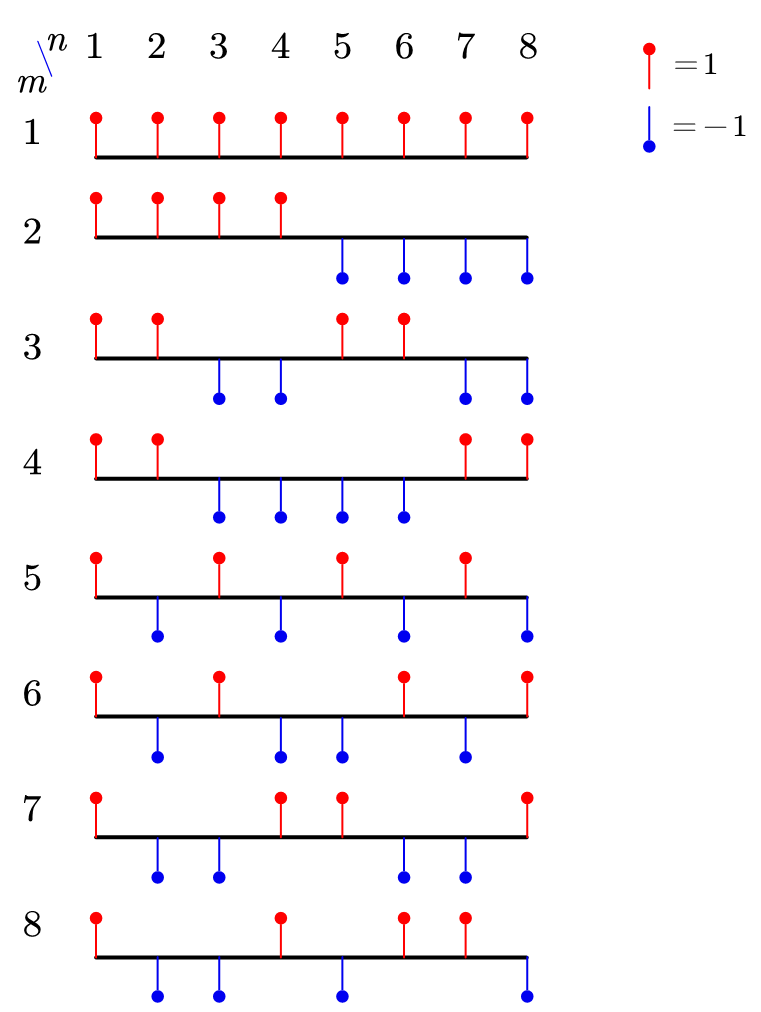}
  \caption{Walsh function \(w(n, \lambda)\), (n = 1, \ldots\, 8)}
  \label{fig:walsh_functions}
\end{figure}

\paragraph{Walsh-Fourier Transformation} The Walsh-Fourier Transform employs the same fundamental idea as the Fourier Transform. However, unlike the Fourier Transform, which uses a system of sines and cosines, the Walsh-Fourier Transform utilizes Walsh functions. These functions consist of square waves with only two values, +1 and -1 \cite{stoffer1991walsh}. Figure~\ref{fig:walsh_functions} illustrates a set of Walsh functions, denoted as \(w(n, \lambda)\) (where \(n = 0, 1, 2, \ldots\) and \(0 \leq \lambda < 1\)). These Walsh functions are distinguished by the number of times they change signs within a unit interval. As Walsh functions are non-periodic, the index \(n\) in \(w(n, \lambda)\) cannot be termed as frequency. Hence, Harmuth and Lee\cite{harmuth1971transmission}introduced the term \textbf{\textit{sequency}} to describe the generalized frequency and to differentiate the functions.

In practical calculations, we utilize the fast version of the Walsh-Fourier Transform (FWFT), proposed by Shanks\cite{shanks1969computation}. As the period of Walsh functions is \(M = 2^n\), forming a complete orthogonal set with \(M\) different functions, if the length \(T\) of the time series \(X\) is not a power of two, then \(X\) must be padded with zeros to become \(X = \{x_{t=i} \mid i = 1, 2, \ldots, N\) and \(t = 1, 2, \ldots, T_2\}\), where \(T_2 = 2^{\lfloor \log_2 T \rfloor + 1}\), and \(\lfloor c \rfloor\) denotes the integer part of \(c\).

The Walsh functions \(wal(m, n)\) for \(m, n = 1, 2, \ldots, T_2\) are defined as follows:
The first two Walsh functions are defined as,
\[
wal(1, n) = 1, \quad n = 1, 2, \ldots, T_2,
\]
\[
wal(2, n) = 
\begin{cases} 
1 & \text{if } n = 1, 2, \ldots, {T_2}/{2} - 1, \\
-1 & \text{if } n = {T_2}/{2}, {T_2}/{2} + 1, \ldots, T_2.
\end{cases}
\]
The rest of the set is generated through the iterative equation,
\[
wal(m, n) = wal\left(\left\lfloor \frac{m}{2} \right\rfloor, 2n\right) \cdot wal\left(m - 2\left\lfloor \frac{m}{2} \right\rfloor, n\right)
\]
According to the Walsh functions, we define the FWFT. Given an array of real numbers \(f(n)\) of length \(T_2\), its Walsh-Fourier transform is given by:
\[
F(m) = \sum_{n=1}^{T_2} f(n) \cdot wal(m, n), \quad m = 1, 2, \ldots, T_2
\]
Correspondingly, the inverse Walsh-Fourier transform is defined as:
\[
f(n) = \frac{1}{T_2} \sum_{m=1}^{T_2} F(m) \cdot wal(n, m), \quad n = 1, 2, \ldots, T_2
\]
Walsh functions are not unique. Different Walsh functions will result in different transform outcomes, and these differences propagate along time, i.e., for larger \(T_2\), the difference in the frequency domain images caused by different Walsh functions becomes more significant. However, for clustering tasks, as long as the same set of Walsh functions is used to represent the frequency domain for all categorized time series in the dataset, the differentiation in the frequency space is maintained.

\paragraph{Persistence homology} Persistent homology is an emerging method in TDA, typically applied to point cloud data or function graphs. Its purpose is to unearth features within data that persist across multiple scales \cite{edelsbrunner2008persistent,pereira2015persistent}. In conjunction with WFT analysis, this study employs persistent homology based on function graphs.

\begin{figure} [H]
  \centering
  \includegraphics[width=0.8\textwidth]{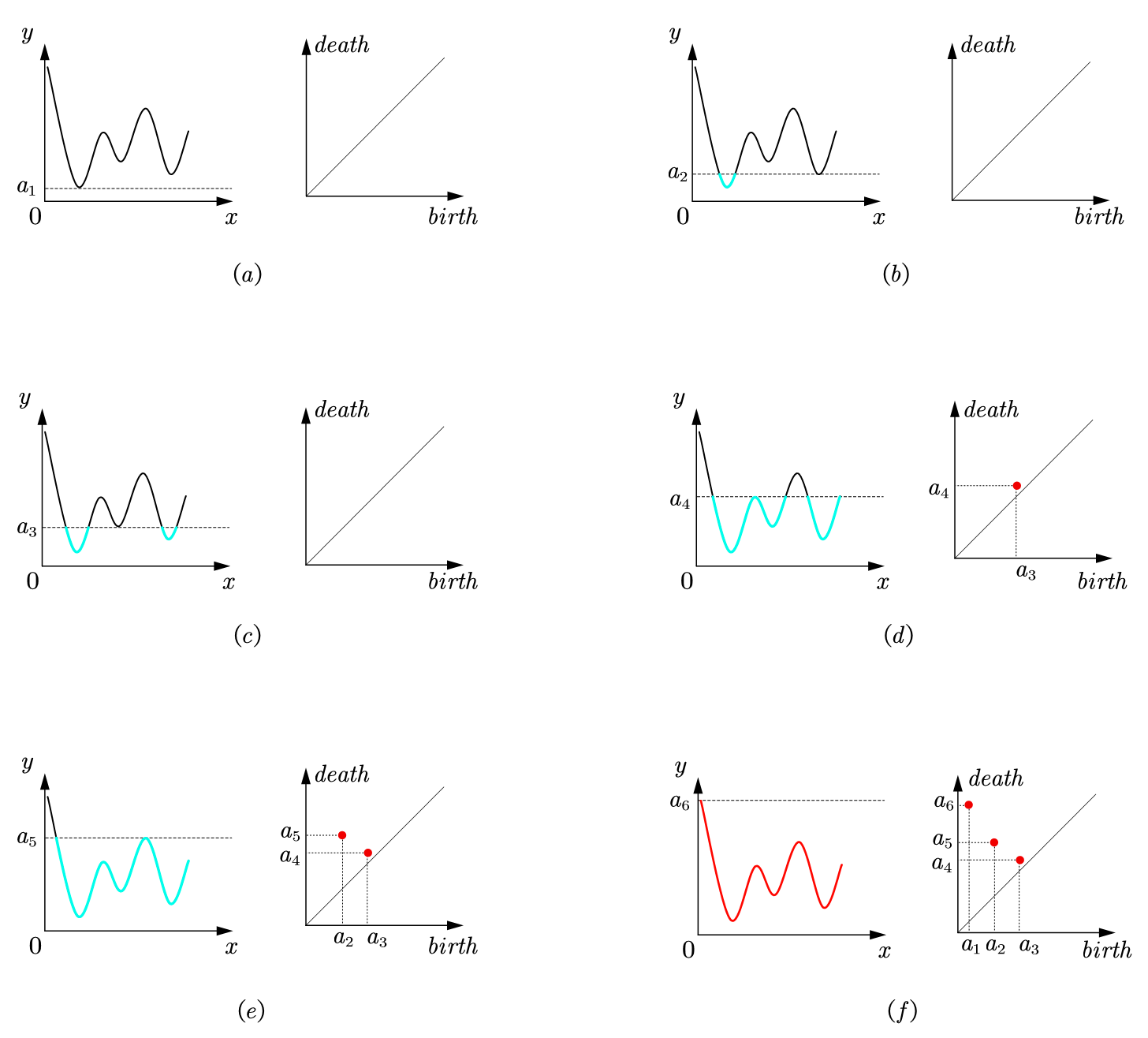}
  \caption{An example: persistence homology using in function graph}
  \label{fig:ph_example}
\end{figure}

Persistent homology used for analyzing function graphs, as illustrated in Figure~\ref{fig:ph_example}. Given a simplicial complex \( K \) with a vertex set \( V \) and a function \( f: V \rightarrow \mathbb{R} \), for any simplex \( \sigma = [v_0, \ldots, v_k] \in K \), define \( f([v_0, \ldots, v_k]) = \max \{f(v_i) : i = 1, \ldots, k\} \). The function \( f \) can be extended to all simplices of \( K \), and the subcomplex \( K_{\epsilon} = \{\sigma \in K : f(\sigma) \leq \epsilon\} \) is defined as the sublevel set filtration of \( f \). Persistent homology tracks various changes in the filtration, such as the emergence of new connected components or the merging of existing ones, and identifies the features along with their lifespans. The resultant information is encoded as a set of intervals known as barcodes, or equivalently, as a series of points in \( \mathbb{R}^2 \), where the coordinates correspond to the start and end times of these intervals. Such a graph is referred to as a \textbf{\textit{birth-death diagram}}.

\begin{figure}
  \centering
  \includegraphics[width=0.8\textwidth]{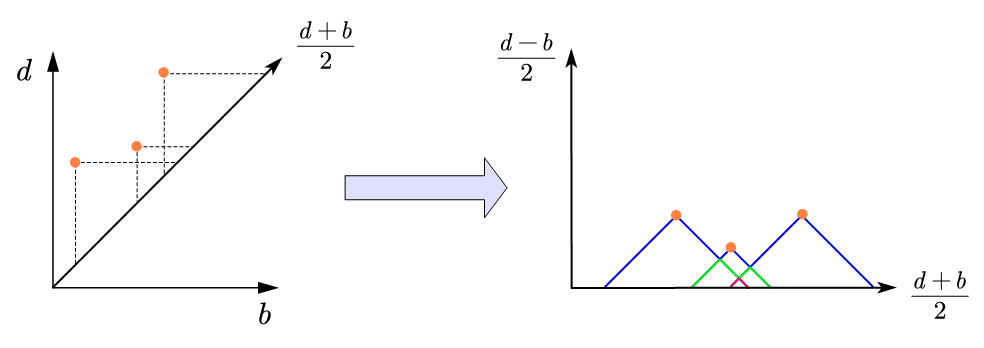}
  \caption{Transformation from birth-death diagram to persistence landscape}
  \label{fig:ph_landscape}
\end{figure}

Bubenik\cite{bubenik2015statistical} introduced the concept of persistent landscapes, which can be seen as an alternative representation of persistence diagrams, but offering quantifiable features (see Figure~\ref{fig:ph_landscape}). This method aims to represent the topological information encoded in the persistence diagrams as elements within a Hilbert space, thereby facilitating the application of statistical learning methods. Persistent landscapes are a collection of continuous piecewise linear functions \( \lambda: \mathbb{N} \times \mathbb{R} \rightarrow \mathbb{R} \), which summarize the persistence diagram, denoted as \( \textit{dgm} \).

As illustrated in Figure~\ref{fig:ph_landscape}, a birth-death pair \( p = (b, d) \in \textit{dgm} \) is transformed into the point \( \left(\frac{d+b}{2}, \frac{d-b}{2}\right) \). Consequently, consider the following set of functions to define the $p$-th level of the persistent landscape, which characterizes the rotated features of the persistence diagram.

\[
\Lambda_p(t) = 
\begin{cases} 
t - b, & \text{if } t \in \left[b, \frac{d+b}{2}\right] \\
d - t, & \text{if } t \in \left[\frac{d+b}{2}, d\right] \\
0, & \text{otherwise}
\end{cases}
\]

The persistent landscape of \( \textit{dgm} \), denoted as \( \lambda_{\textit{dgm}} \), is obtained by superimposing the graphs of the functions \( \{ \Lambda_p \}_{p \in \textit{dgm}} \) which form an array of piecewise linear curves. Formally, \( \lambda_{\textit{dgm}} \) is the collection of functions \( \lambda_{\textit{dgm}}(k, t) \):

\[
\lambda_{\text{dgm}}(k, t) = k\max_{r \in \text{dgm}} \Lambda_r(t), \quad t \in [0, T], k \in \mathbb{N}
\]

where \( k\max \) denotes the $k$-th largest value in the set, e.g., \( 1\max \) represents the commonly used maximum value function. For a given \( k \in \mathbb{N} \), the function \( \lambda_{\textit{dgm}}(k, \cdot) : \mathbb{R} \rightarrow \mathbb{R} \) is called the $k$-th landscape of \( \textit{dgm} \).

\begin{figure}
  \centering
  \includegraphics[width=0.8\textwidth]{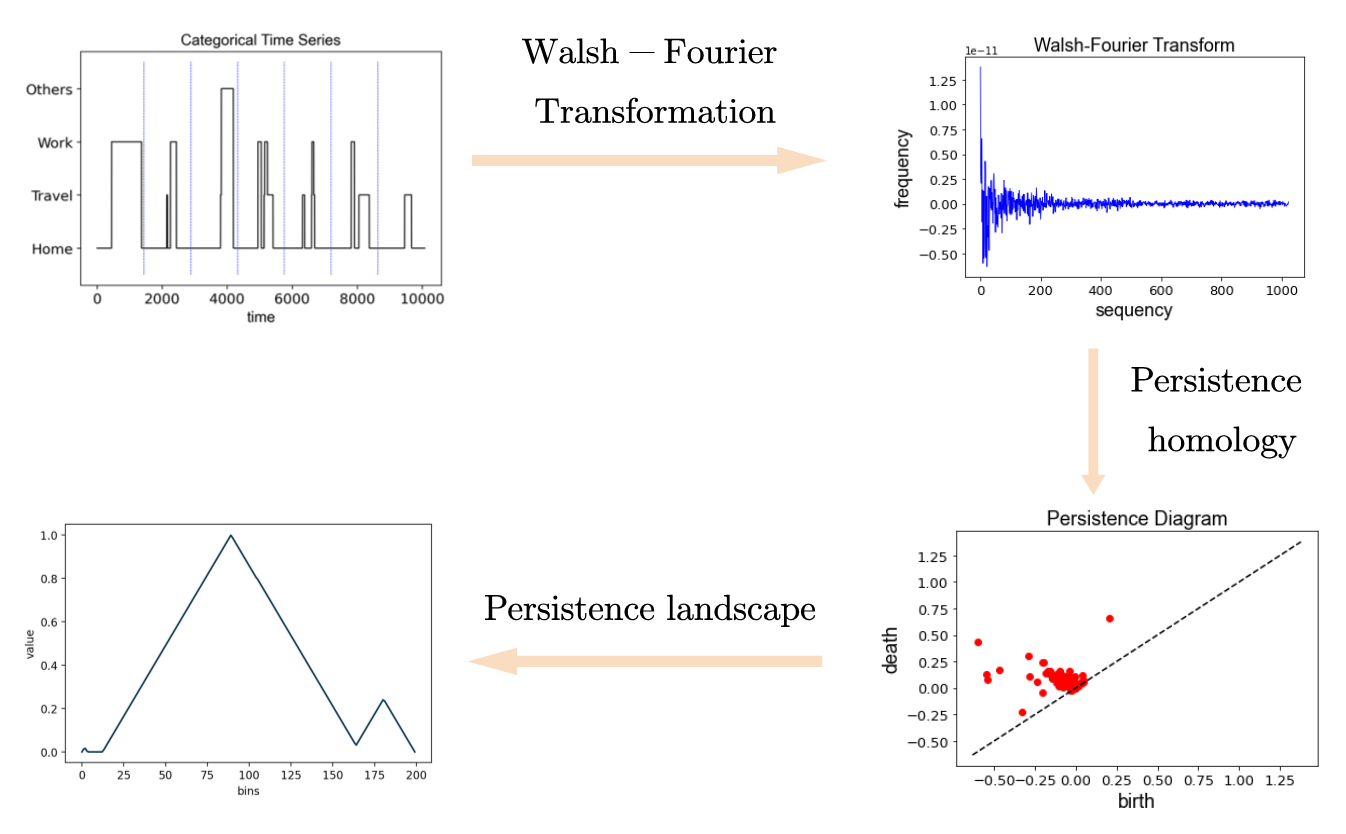}
  \caption{Topological data analysis for mining activity travel pattern}
  \label{fig:TDA_for_pattern}
\end{figure}

Figure~\ref{fig:TDA_for_pattern} shows an example of extracting topological features from an activity-travel sequence. It should be noted that this is merely a brief description of the implementation algorithms for persistent homology and landscapes. The mathematical theory behind persistent homology extends beyond the scope of this article. Interested readers are referred to \cite{bubenik2015statistical,chazal2021introduction,edelsbrunner2008persistent}.

\subsubsection{Geometric measurement of activity-travel sequences}
The application of TDA has provided a new perspective for analyzing categorical time series of activity-travel patterns. However, as a global method, TDA can only extract global information from the time series and fails to consider local quantitative information\cite{carlsson2009topology}. Zhang et al.\cite{zhang2021time} argue that a reasonable and effective framework for time series clustering should consider both global and local information; therefore, this study has also developed geometric methods as a complement to TDA. The essence of the geometric metric for activity-travel sequences is the measurement of sequence similarity. In this study, we adopted methods that weigh the distances measured by edit distance and agenda dissimilarity\cite{allahviranloo2017modeling}.

\paragraph{Sequence alignment method and edit distance} The edit distance is generated by the sequence alignment method (SAM), which has been widely used in the biomedical field since the 1980s, such as comparing the similarity of DNA sequences\cite{corpet1988multiple} or more recent protein sequences. The fundamental idea of SAM is to precisely define the cost metric of the distance between two possible states at specific positions within a sequence. Based on this metric, the pairwise distance between two sequences is defined as the minimum cost of transforming one sequence into another through substitutions, insertions, or deletions. The most commonly used distance is the edit distance, also known as the Levenshtein distance.

Suppose there are two sequences \(x\) and \(y\) with lengths \(L_x\) and \(L_y\) respectively. Based on the Needleman-Wunsch algorithm\cite{needleman1970general}, the edit distance between \(x\) and \(y\) is obtained in the following two steps:

\begin{enumerate}
    \item \textbf{Initialize the distance matrix \(D_{L_x+1, L_y+1}\), let}:
    \[
    D(0, 0) = 0
    \]
    \[
    D(i, 0) = i \times d, \quad \text{for } i = 0, \ldots, L_x
    \]
    \[
    D(0, j) = j \times d, \quad \text{for } j = 0, \ldots, L_y
    \]

    \item \textbf{Iterate to compute minimum costs}:
    \[
    D(i, j) = \min
    \begin{cases}
    D(i-1, j) + d & \text{(deletion)} \\
    D(i, j-1) + d & \text{(insertion)} \\
    D(i-1, j-1) + P(x(i), y(j)) & \text{(substitution)}
    \end{cases}
    \]
    where \(d\) represents the penalty for insertion and deletion, and \(P(x(i), y(j))\) is the cost of substituting \(x(i)\) with \(y(j)\).
\end{enumerate}
Here, \(d\) represents the penalty associated with insertions and deletions, set to 1 in this context. \(P\) is the distance matrix related to substitutions, defined as:
\[
P(x(i), y(j)) = 
\begin{cases} 
1, & \text{if } x(i) \neq y(j) \\
0, & \text{if } x(i) = y(j)
\end{cases}
\]
After completing the iterations, the edit distance \(M_1\) between sequences \(x\) and \(y\) is given by \(M_1 = D(L_x + 1, L_y + 1)\). The result of the edit distance depends on the penalty parameters \(d\) and the substitution cost matrix \(P\).

\begin{figure}
  \centering
  \includegraphics[width=0.8\textwidth]{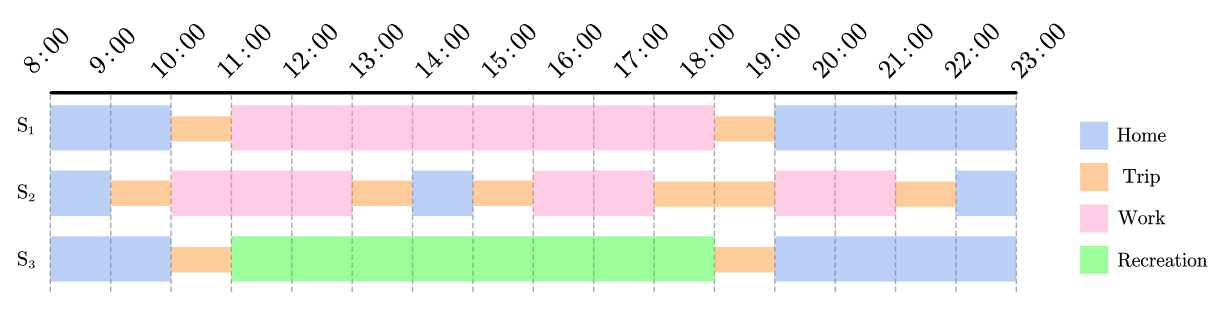}
  \caption{The activity travel sequences of samples \(S_1\), \(S_2\), \(S_3\) (with an hour interval)}
  \label{fig:activity_travel_sequences}
\end{figure}

\paragraph{Agenda dissimilarity} Relying only on the SAM is insufficient for effectively distinguishing between activity patterns. To illustrate this, consider the example given in Figure~\ref{fig:activity_travel_sequences} and Table~\ref{tab:comparison_matrix}, where the edit distance between \( S_1 \) and \( S_2 \) is 8, which equals the distance between \( S_1 \) and \( S_3 \). However, both \( S_1 \) and \( S_2 \) involve home and work activities, differing only in timing, whereas \( S_3 \) includes a significant period of recreational activities outside, indicating an agenda dissimilarity. Nevertheless, in SAM, \( S_2 \) and \( S_3 \) appear same to \( S_1 \), which is clearly unreasonable. Therefore, we utilize agenda dissimilarity as a second metric to complement the sequence alignment method.

Agenda dissimilarity, based on the content of sequences \( A \) and \( B \), is defined as follows:

\[
M_2 = Y \left(1 - \frac{|A \cap B|}{|A \cup B|}\right)
\]

where \( Y \) is the agenda parameter, set to match the maximum possible value of the edit distance \( M_1 \), ensuring that both metrics are within a comparable range. In the example shown in Figure 9, \( Y \) is set to 16. From the definition, it is known that \( M_2 \) has only two possible values. If sequences \( A \) and \( B \) contain the same types of activities, then \( M_2 = 0 \); if \( A \) and \( B \) contain more than one differing type of activity, then \( M_2 = Y \). 

The overall similarity measure \( M \) between sequences is determined by the weighted sum of \( M_1 \) and \( M_2 \):\[M = w_1 \cdot M_1 + w_2 \cdot M_2\] It should be noted that different values of the weights \( w_1 \) and \( w_2 \) can significantly affect the clustering outcomes.

\subsubsection{Pattern clustering}
Persistent landscapes (topological features) and sequence similarity (geometric features) are used as inputs for a two-stage clustering algorithm called CAK means, which combines Affinity Propagation and K-means clustering \cite{zhu2009initializing}. The fundamental idea behind CAK means clustering is to use Affinity Propagation to initialize the centroids and the number of clusters for K-means clustering, conducted in two steps.

First, Affinity Propagation is applied to determine the quantity and locations of initial centroids within the data, eliminating the arbitrariness of presetting parameters inherent in K-means. In the second step, the results from Affinity Propagation (initial centroids and cluster counts) are used as inputs to implement K-means clustering.

CAK means has several advantages; compared to traditional K-means clustering, it requires less computation time and results in smaller classification errors. It evaluates each data point as a potential centroid and generates the optimal number of clusters from the data itself, rather than setting hyperparameters externally.

\begin{table}
\centering
\caption{Comparison of Edit Distance and Agenda Dissimilarity for Sequences S1, S2, S3}
\label{tab:comparison_matrix}
\begin{tabular}{ccccccccc}
\toprule
& \multicolumn{3}{c}{Edit Distance} && \multicolumn{3}{c}{Agenda Dissimilarity} \\
\cmidrule{2-4} \cmidrule{6-8}
& S1 & S2 & S3 && S1 & S2 & S3 \\
\midrule
S1 & 0 & 8 & 8 && 0 & 0 & 16 \\
S2 & 8 & 0 & 13 && 0 & 0 & 16 \\
S3 & 8 & 13 & 0 && 16 & 16 & 0 \\
\bottomrule
\end{tabular}
\end{table}

\begin{figure} [h]
  \centering
  \includegraphics[width=0.8\textwidth]{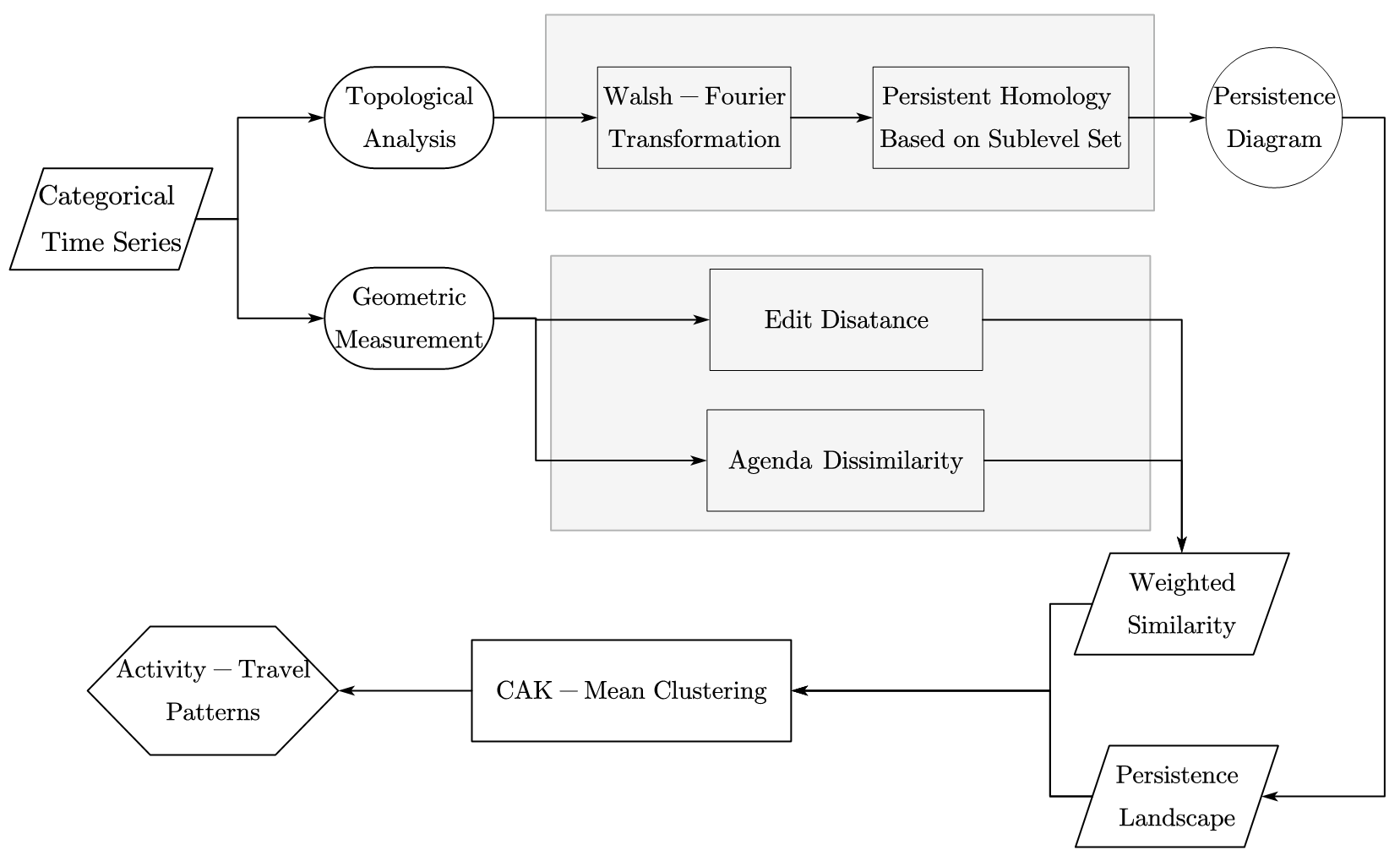}
  \caption{Activity-travel pattern mining workflow}
  \label{fig:minning_workflow}
\end{figure}

In summary, the pattern mining framework adopted in this study is illustrated in Figure~\ref{fig:minning_workflow}.

\section{Results}
\subsection{Implementation of clustering algorithm}
\subsubsection{Optimal hyperparameters of clustering}
\begin{figure} [H]
  \centering
  \begin{minipage}{0.45\textwidth}
    \centering
    \includegraphics[width=\textwidth]{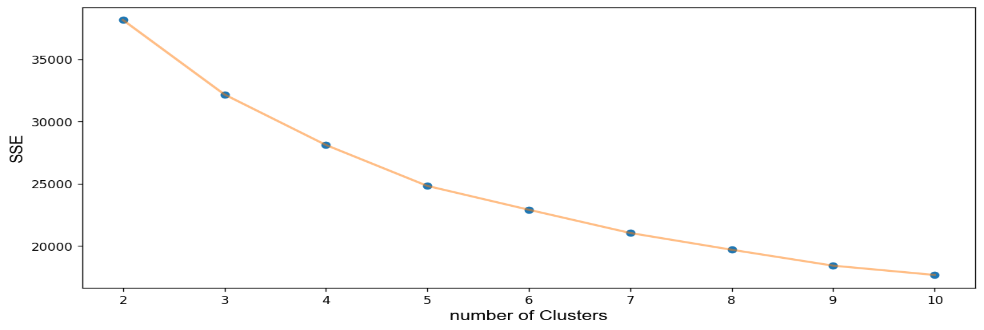}
    \caption{Clustering elbow plot}
    \label{fig:elbow_plot}
  \end{minipage}%
  \begin{minipage}{0.45\textwidth}
    \centering
    \includegraphics[width=\textwidth]{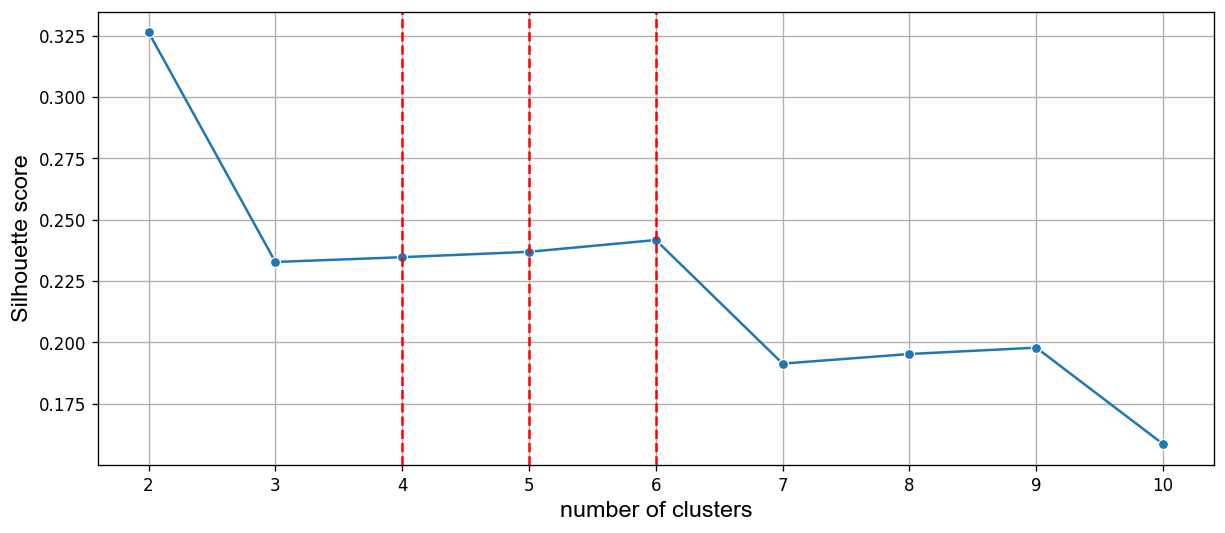}
    \caption{Clustering Silhouette Coefficient Plot}
    \label{fig:sihouette_plot}
  \end{minipage}
\end{figure}

This study employs the elbow method and silhouette coefficient as criteria for determining the optimal number of clusters in K-means clustering. The elbow method uses the sum of squared errors (SSE) as a metric, which represents the quality of clustering. The SSE is calculated as follows:
\[
\textit{SSE} = \sum_{i=1}^{k} \sum_{p \in C_i} \|p - m_i\|^2
\]
where \( p \) represents a point in the dataset, and \( m_i \) is the centroid of the cluster \( C_i \). In cluster analysis, the elbow method determines the best number of clusters by observing changes in the total SSE as the number of clusters \( k \) increases. When \( k \) is less than the true number of clusters, increasing \( k \) significantly reduces the SSE; however, as \( k \) surpasses the true number of clusters, the rate of decrease in SSE slows down and tends to stabilize, forming an "elbow" shape (see Figure~\ref{fig:elbow_plot}). The \( k \) value corresponding to this elbow point is considered the optimal number of clusters.

Another performance metric, the silhouette coefficient, ranges from [-1, 1], with higher values indicating better clustering performance. Specifically, the silhouette coefficient \(s(i)\) for a sample \(i\) in the dataset is defined as follows:

\[
s(i) = \frac{b(i) - a(i)}{\max(a(i), b(i))}
\]

where \(a(i)\) represents the average distance between sample \(i\) and all other samples in its cluster, and \(b(i)\) represents the average distance between sample \(i\) and all samples in the nearest cluster to its own. The silhouette coefficient for the entire dataset is the average of all the silhouette coefficients for all points in the dataset. Figure~\ref{fig:sihouette_plot} illustrates the variation of the silhouette coefficient with the number of clusters \(k\) in this experiment. Combining the SSE and the silhouette coefficient as metrics, we can determine that the optimal number of clusters is 6.

\subsubsection{Algorithm performance}
\begin{figure}
  \centering
  \includegraphics[width=0.6\textwidth]{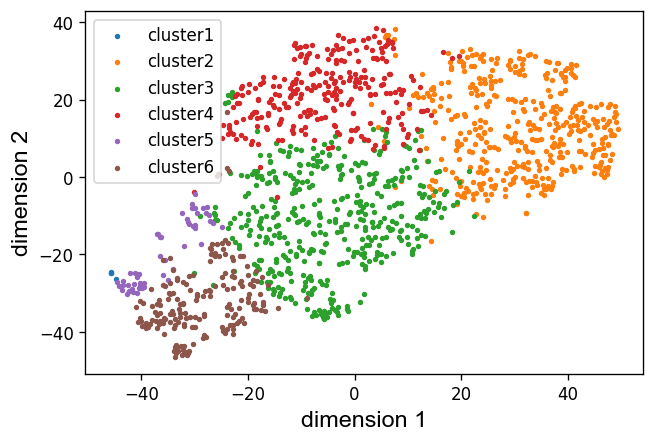}
  \caption{Clustering results using t-SNE dimensionality reduction}
  \label{fig:tsne}
\end{figure}

We utilize t-SNE (t-Distributed Stochastic Neighbor Embedding) for data dimensionality reduction to observe the clustering effects. t-SNE works by converting the similarities between data points into joint probabilities and aims to minimize the discrepancy between the probabilities in high-dimensional and low-dimensional spaces. Although t-SNE cannot replace methods like PCA or SVD, it is suitable for visualizing the clustering results of this experiment. Figure~\ref{fig:tsne} shows the t-SNE clustering effects with different categories distinguished by different colors, from which it can be seen that the clustering results are ideal.

Additionally, while analyzing the clustering results, it was found that the first cluster contained only 4 samples, and the fifth cluster contained only 52 samples. Since these two clusters are the closest in the t-SNE plot, we decided to merge the first and fifth clusters, ultimately obtaining five categories of activity-travel patterns.

This experiment also explored other classic clustering algorithms, including K-means, K-medoids, and fuzzy C-means, and used the Caliński-Harabasz (CH) score to evaluate the clustering effectiveness of different algorithms. The CH score considers both the compactness within clusters and the separation between clusters, with a higher score indicating better clustering performance \cite{calinski1974dendrite}. The CH scores for each clustering algorithm are shown in Figure~\ref{}, where it is evident that CAK means outperforms all other methods.

\begin{figure}
  \centering
  \includegraphics[width=0.6\textwidth]{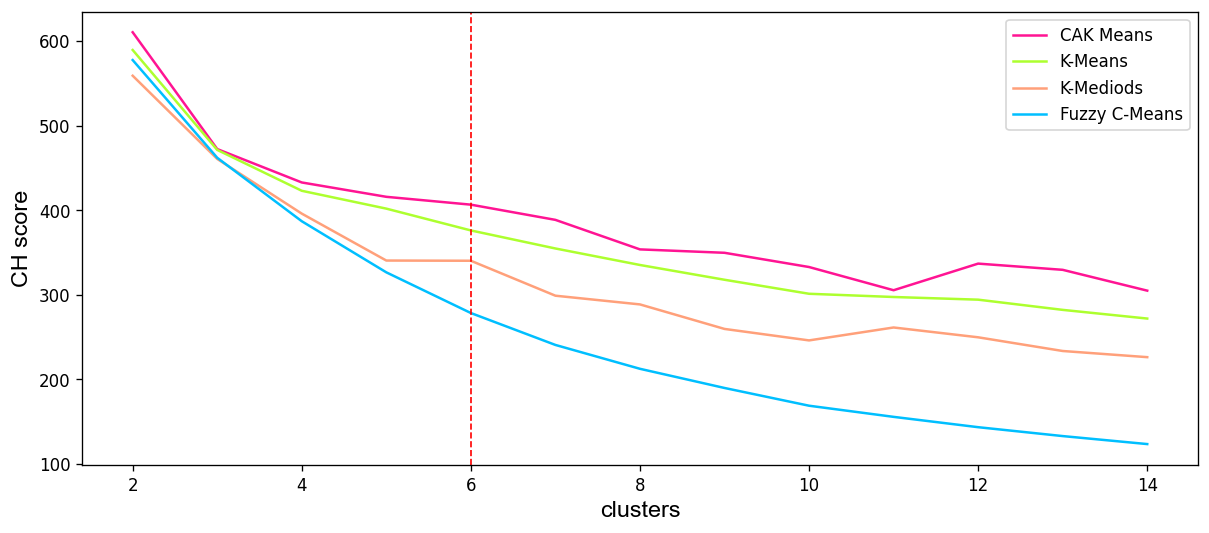}
  \caption{Comparison of CH scores among different clustering algorithms}
  \label{fig:ch_score}
\end{figure}

\subsection{Results of activity-travel patterns}
\begin{figure}
  \centering
  \includegraphics[width=0.8\textwidth]{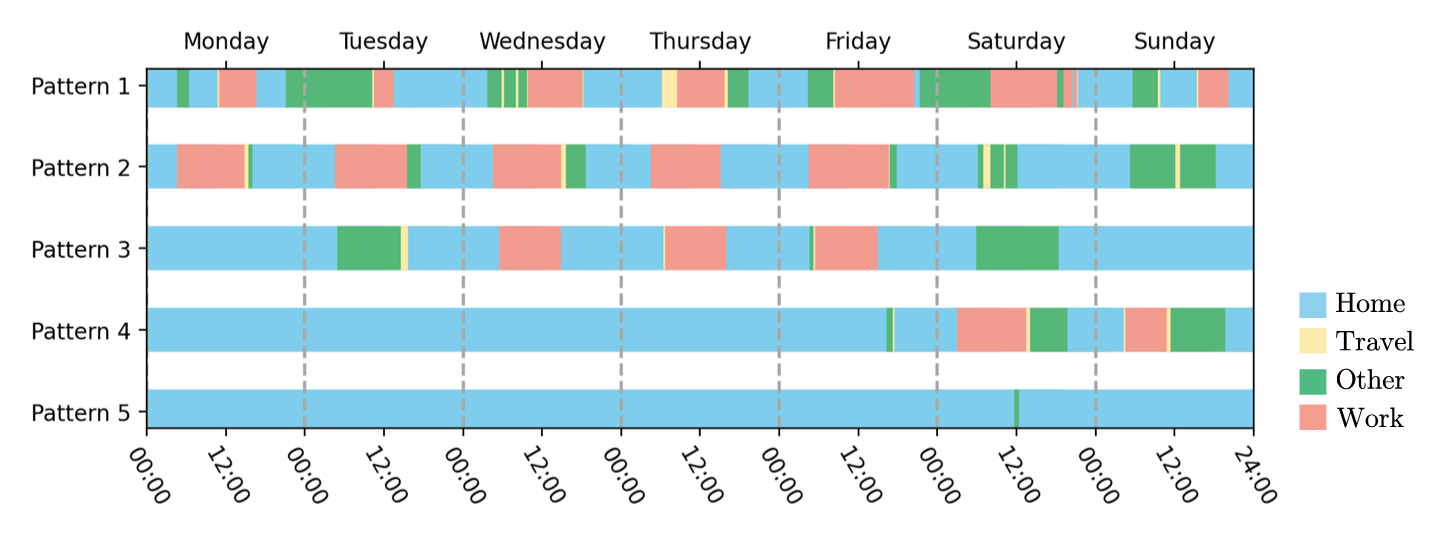}
  \caption{Representative Weekly Sequence of Activity and Travel Time Utilization Patterns}
  \label{fig:representative_pattern}
\end{figure}

The clustering resulted in five groups of categorized time series, representing residents with five different life rhythms in the community. Figure~\ref{fig:representative_pattern} shows the representative sequence of each category, which is the data represented by the cluster centroid.

Visually, we can easily distinguish that the time use pattern of Pattern 1 has the most frequent and complex travel chains, so we call these residents \textit{multitasking}. In contrast, residents of Pattern 5 spend most of their time at home, with the least periodicity in travel (\textit{home activities-dominant}). Residents in Pattern 2 and Pattern 3 have very similar life rhythms, but residents in Pattern 2 almost regularly return home after work every day, making it more periodic than Pattern 3. Therefore, we name Pattern 2 residents as \textit{work-dominant}. Residents in Pattern 3 work less frequently than those in Pattern 2, allowing more time for other activities. Hence, we consider these people to have a more balanced distribution of time among work, other activities, and home activities, belonging to the \textit{balanced} category. The last group, Pattern 4, displays a time use pattern that deviates from the norm. As shown in Figure 16, residents in Pattern 4 choose to stay home during weekdays, which might include activities such as remote work, homeschooling, and household chores. It is clear that their time use pattern does not follow the traditional "work on weekdays, rest on weekends" rhythm but rather reverses it. Therefore, we define this group as \textit{reverse rhythm}.

\begin{figure} [H]
  \centering
  \includegraphics[width=0.8\textwidth]{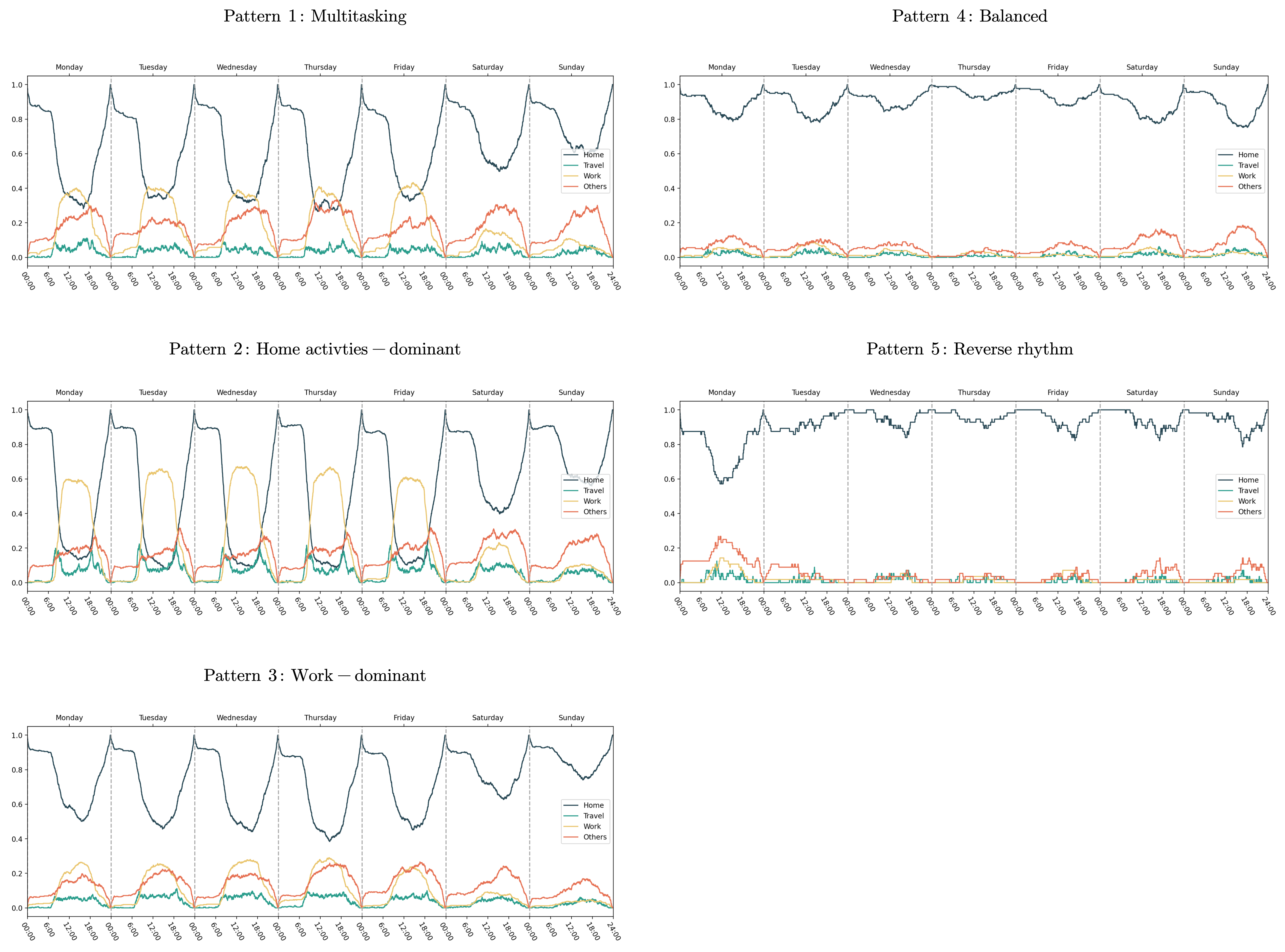}
  \caption{Proportions of Activity Types Across Five Patterns}
  \label{fig:proportions_activity}
\end{figure}

To facilitate comparison of each group's characteristics, Figure~\ref{fig:proportions_activity} shows the proportion of activity types for each category of residents by hour throughout the week. Additionally, Table~\ref{tab:cluster_statistics} presents the socioeconomic statistics of each category of residents. From the table, it can be seen that the \textit{multitasking} residents have employment rates and monthly communication fees far exceeding the average, reflecting the busiest work-life rhythm. The \textit{work-dominant} residents show no gender differences from the average, but only 0.8(\%) of the elderly (over 65 years) are in this category. Residents in Pattern 3, the \textit{balanced} type, tend to be younger, primarily individuals under 35 years old. These individuals have a higher demand for other activities, such as entertainment and shopping, while also having compulsory demands like work (with 74.8(\%) being employed) and study. Therefore, Pattern 3 shows the most balanced proportion of work and other activities among all patterns. The \textit{reverse rhythm} residents are more represented in the 16-24 and 50-59 age groups, such as university students and newly retired individuals. The commonality between these age groups is the lack of compulsory work requirements but good physical conditions. This also explains why their life rhythm differs from the norm, as they have greater discretionary control over their time. Lastly, the "home activities-dominant" residents show a female proportion about 6(\%) higher than the average, with a higher proportion of middle-aged and elderly individuals, and only 53.6(\%) are employed. It can be inferred that this category predominantly comprises retired women.

\begin{table}[H]
\centering
\caption{Demographic Statistics by Clusters}
\label{tab:cluster_statistics}
\begin{tabular}{llcccccc}
\toprule
Variable & Category & Overall (\%) & \multicolumn{5}{c}{Proportion within Pattern Clusters (\%)} \\
\cmidrule(lr){4-8}
& & & 1 (c1) & 2 (c3) & 3 (c2) & 4 (c5) & 5 (c0) \\
\midrule
Gender & Male & 72.4 & 70.9 & 70.7 & 71.5 & 69.7 & 66.1 \\
& Female & 27.6 & 29.1 & 29.3 & 28.5 & 30.3 & 33.9 \\
Age & 13-15(3) & 0.5 & 0 & 0.8 & 0.5 & 0 & 0 \\
& 16-18(4) & 1.6 & 0 & 1.5 & 3.1 & 2.8 & 1.8 \\
& 19-24(5) & 15.1 & 8.2 & 13.1 & 15.2 & 21.9 & 19.6 \\
& 25-29(6) & 11.5 & 10.0 & 18.1 & 11.4 & 7.3 & 7.1 \\
& 30-34(7) & 23.4 & 25.9 & 23.2 & 18.5 & 11.2 & 23.2 \\
& 35-39(8) & 20.8 & 24.9 & 19.3 & 24.0 & 19.7 & 16.1 \\
& 40-44(9) & 7.8 & 6.7 & 8.9 & 6.7 & 8.4 & 7.1 \\
& 45-49(10) & 4.7 & 6.2 & 2.7 & 2.6 & 7.9 & 7.1 \\
& 50-54(11) & 7.8 & 8.2 & 7.3 & 7.8 & 9.6 & 8.9 \\
& 55-59(12) & 4.2 & 5.5 & 4.2 & 7.1 & 7.3 & 5.4 \\
& 65-69(14) & 1.6 & 4.5 & 0.4 & 1.7 & 2.8 & 3.6 \\
& 70+(15) & 1.0 & 0 & 0.4 & 1.4 & 1.1 & 0 \\
Employment & Employed & 76.6 & 90.5 & 81.1 & 74.8 & 52.2 & 53.6 \\
Mobile Brand & Apple & 39.1 & 38.6 & 37.8 & 35.6 & 38.2 & 35.7 \\
& Huawei & 31.8 & 40.8 & 35.9 & 36.3 & 24.7 & 21.4 \\
& Xiaomi & 7.8 & 10.4 & 3.9 & 6.4 & 10.1 & 7.1 \\
& Others & 21.3 & 10.2 & 22.4 & 21.7 & 27.0 & 35.8 \\
\midrule
Variable & Metric & Average & \multicolumn{5}{c}{Weekly Pattern Clusters Statistics} \\
\cmidrule(lr){4-8}
& & & 1 (c1) & 2 (c3) & 3 (c2) & 4 (c5) & 5 (c0) \\
\midrule
Monthly Communication Fees & Mean & 87.1 & 111.6 & 90.0 & 91.5 & 71.5 & 77.6 \\
& Std Dev & 71.3 & 89.3 & 69.3 & 80.9 & 52.3 & 73.4 \\
\bottomrule
\end{tabular}
\end{table}

\section{Discussion and Conclusion}
Identifying and summarizing the patterns of complex activity-travel behaviors among urban residents is a crucial prerequisite for exploring urban mobility and achieving precise planning. This study investigates the activity-travel patterns of community residents in Shenzhen based on mobile signaling data. We transformed each user's records into a categorized time series represented by activity types and travel, using Walsh-Fourier transform and topological data analysis to obtain their frequency domain characteristics. These features, combined with geometric features, served as inputs for the clustering algorithm to cluster the time series. The analysis results divided the community residents into five categories, each with different life rhythms.

The conclusions of this study can guide more refined traffic planning policy recommendations and interventions. For example, for multitasking residents, who may engage in various activities during their trips, flexible and convenient transportation methods are needed. Policy recommendations might include providing customized bus services and offering free public Wifi so they can work or perform other activities during their trips. For work-dominant residents, whose transportation needs are often concentrated during peak commuting hours, policy suggestions might include providing high-frequency public transportation services, implementing flexible working hours to disperse peak period pressure, or offering shared bicycles or walking facilities to encourage green commuting for the last mile. For balanced residents, whose travel needs may be more dispersed, around-the-clock transportation services are needed, as well as improved pedestrian and slow-traffic facilities to allow them to travel as needed. For reverse rhythm residents, sufficient transportation services may be needed during weekends or off-peak hours on weekdays. Policy recommendations could include providing off-peak public transportation services or offering night transportation services.

The results of this study may have the following potential applications or practical implications: (1) Due to the existence of activity-travel patterns, planners have the conditions to conduct research on people's travel behaviors on a larger scale and obtain new behavioral evidence. Behavioral evidence can be used to increase transportation options and improve community facilities for target populations\cite{yin2021exploring}; (2) The methods in this paper can also be used to compare activity-travel patterns of residents under different cultures, economic conditions, policies, and periods. For example, analyzing the differences in travel patterns between residents of developed and developing countries can reveal the impact of economic development levels and transportation infrastructure on residents' travel choices. Moreover, comparing changes before and after policy interventions, such as new traffic regulations or urban planning policies, can help evaluate the effectiveness and impact scope of these policies; (3) Residents' life rhythms can serve as a way to reveal and reflect the urban pulse\cite{reades2016finding}, which is key to understanding urban dynamics.

There are some limitations in the data and methods used in this analysis. Firstly, the analysis is based on one month's activity-travel data of residents in one community, which limits the spatial and temporal breadth of the analysis. Expanding the data range to the city level would help to gain a deeper understanding of urban life rhythms and residents' quality of life. Additionally, the activity categories in the dataset used may contain errors, affecting the accuracy of the results. On the other hand, mobile signaling data lack rich individual socioeconomic attribute information, limiting the in-depth interpretation of the results. Methodologically, the current methods used still have high computational costs, limiting large-scale data processing, and the clustering methods lack interpretability, making it difficult to intuitively understand the reasons behind the classifications and results. These issues need to be improved and explored in future research.

\bibliographystyle{unsrt}  
\bibliography{references}

\begin{thebibliography}{10}

\bibitem{golob2003structural}
T.~F. Golob.
\newblock Structural equation modeling for travel behavior research.
\newblock {\em Transportation Research Part B: Methodological}, 37(1):1--25, 2003.

\bibitem{gotschi2017towards}
T.~Götschi, A.~de~Nazelle, C.~Brand, and R.~Gerike.
\newblock Towards a comprehensive conceptual framework of active travel behavior: a review and synthesis of published frameworks.
\newblock {\em Current Environmental Health Reports}, 4:286--295, 2017.

\bibitem{kitamura1988evaluation}
R.~Kitamura.
\newblock An evaluation of activity-based travel analysis.
\newblock {\em Transportation}, 15:9--34, 1988.

\bibitem{dong2006moving}
X.~Dong, M.~E. Ben-Akiva, J.~L. Bowman, and J.~L. Walker.
\newblock Moving from trip-based to activity-based measures of accessibility.
\newblock {\em Transportation Research Part A: Policy and Practice}, 40(2):163--180, 2006.

\bibitem{liu2021similar}
L.~Liu, E.~A. Silva, and Z.~Yang.
\newblock Similar outcomes, different paths: Tracing the relationship between neighborhood-scale built environment and travel behavior using activity-based modelling.
\newblock {\em Cities}, 110:103061, 2021.

\bibitem{shiftan2008use}
Y.~Shiftan.
\newblock The use of activity-based modeling to analyze the effect of land-use policies on travel behavior.
\newblock {\em The Annals of Regional Science}, 42:79--97, 2008.

\bibitem{huang2018tracking}
J.~Huang, D.~Levinson, J.~Wang, J.~Zhou, and Z.-j. Wang.
\newblock Tracking job and housing dynamics with smartcard data.
\newblock {\em Proceedings of the National Academy of Sciences}, 115(50):12710--12715, 2018.

\bibitem{reades2016finding}
J.~Reades, C.~Zhong, E.~Manley, R.~Milton, and M.~Batty.
\newblock Finding pearls in london's oysters.
\newblock {\em Built Environment}, 42(3):365--381, 2016.

\bibitem{zhang2017detecting}
W.~Zhang and J.-C. Thill.
\newblock Detecting and visualizing cohesive activity-travel patterns: A network analysis approach.
\newblock {\em Computers, Environment and Urban Systems}, 66:117--129, 2017.

\bibitem{yang2001infominer}
J.~Yang, W.~Wang, and P.~S. Yu.
\newblock Infominer: Mining surprising periodic patterns.
\newblock In {\em Proceedings of the seventh ACM SIGKDD international conference on Knowledge discovery and data mining}, 2001.

\bibitem{yin2015mining}
F.~Yin, X.~Li, C.~Yao, and L.~Shen.
\newblock Mining frequent spatio-temporal items in trajectory data.
\newblock {\em International Journal of Database Theory and Application}, 8(4):149--156, 2015.

\bibitem{kalnis2005discovering}
P.~Kalnis, N.~Mamoulis, and S.~Bakiras.
\newblock On discovering moving clusters in spatio-temporal data.
\newblock In {\em Advances in Spatial and Temporal Databases: 9th International Symposium, SSTD 2005, Angra dos Reis, Brazil, August 22-24, 2005. Proceedings}, volume~9, 2005.

\bibitem{giannotti2007trajectory}
F.~Giannotti, M.~Nanni, F.~Pinelli, and D.~Pedreschi.
\newblock Trajectory pattern mining.
\newblock In {\em Proceedings of the 13th ACM SIGKDD International Conference on Knowledge Discovery and Data Mining}, 2007.

\bibitem{mamoulis2004mining}
N.~Mamoulis, H.~Cao, G.~Kollios, M.~Hadjieleftheriou, Y.~Tao, and D.~W. Cheung.
\newblock Mining, indexing, and querying historical spatiotemporal data.
\newblock In {\em Proceedings of the tenth ACM SIGKDD international conference on Knowledge discovery and data mining}, 2004.

\bibitem{li2016probabilistic}
J.~Li, J.~Wang, J.~Zhang, Q.~Qin, T.~Jindal, and J.~Han.
\newblock A probabilistic approach to detect mixed periodic patterns from moving object data.
\newblock {\em GeoInformatica}, 20:715--739, 2016.

\bibitem{yuan2017pred}
Q.~Yuan, W.~Zhang, C.~Zhang, X.~Geng, G.~Cong, and J.~Han.
\newblock Pred: Periodic region detection for mobility modeling of social media users.
\newblock In {\em Proceedings of the Tenth ACM International Conference on Web Search and Data Mining}, 2017.

\bibitem{bermingham2020mining}
L.~Bermingham and I.~Lee.
\newblock Mining distinct and contiguous sequential patterns from large vehicle trajectories.
\newblock {\em Knowledge-Based Systems}, 189:105076, 2020.

\bibitem{zhang2018hierarchical}
D.~Zhang, K.~Lee, and I.~Lee.
\newblock Hierarchical trajectory clustering for spatio-temporal periodic pattern mining.
\newblock {\em Expert Systems with Applications}, 92:1--11, 2018.

\bibitem{zhang2019mining}
D.~Zhang, K.~Lee, and I.~Lee.
\newblock Mining hierarchical semantic periodic patterns from gps-collected spatio-temporal trajectories.
\newblock {\em Expert Systems with Applications}, 122:85--101, 2019.

\bibitem{shi2019mining}
T.~Shi, G.~Ji, Y.~Liu, and B.~Zhao.
\newblock Mining group periodic moving patterns from spatio-temporal trajectories.
\newblock In {\em 2019 Seventh International Conference on Advanced Cloud and Big Data (CBD)}, 2019.

\bibitem{chazal2021introduction}
F.~Chazal and B.~Michel.
\newblock An introduction to topological data analysis: fundamental and practical aspects for data scientists.
\newblock {\em Frontiers in Artificial Intelligence}, 4:108, 2021.

\bibitem{lum2013extracting}
P.~Y. Lum, G.~Singh, A.~Lehman, T.~Ishkanov, M.~Vejdemo-Johansson, M.~Alagappan, J.~Carlsson, and G.~Carlsson.
\newblock Extracting insights from the shape of complex data using topology.
\newblock {\em Scientific Reports}, 3(1):1--8, 2013.

\bibitem{edelsbrunner2008persistent}
H.~Edelsbrunner and J.~Harer.
\newblock Persistent homology-a survey.
\newblock {\em Contemporary Mathematics}, 453:257--282, 2008.

\bibitem{feng2020spatial}
M.~Feng and M.~A. Porter.
\newblock Spatial applications of topological data analysis: Cities, snowflakes, random structures, and spiders spinning under the influence.
\newblock {\em Physical Review Research}, 2(3):033426, 2020.

\bibitem{ravishanker2021introduction}
N.~Ravishanker and R.~Chen.
\newblock An introduction to persistent homology for time series.
\newblock {\em Wiley Interdisciplinary Reviews: Computational Statistics}, 13(3):e1548, 2021.

\bibitem{wu2017congestion}
Y.~Wu, G.~Shindnes, V.~Karve, D.~Yager, D.~B. Work, A.~Chakraborty, and R.~B. Sowers.
\newblock Congestion barcodes: Exploring the topology of urban congestion using persistent homology.
\newblock In {\em 2017 IEEE 20th International Conference on Intelligent Transportation Systems (ITSC)}, 2017.

\bibitem{yang2021detecting}
Y.~Yang, C.~Xiong, J.~Zhuo, and M.~Cai.
\newblock Detecting home and work locations from mobile phone cellular signaling data.
\newblock {\em Mobile Information Systems}, 2021:1--13, 2021.

\bibitem{millward2019activity}
H.~Millward, M.~H. Hafezi, and N.~S. Daisy.
\newblock Activity travel of population segments grouped by daily time-use: Gps tracking in halifax, canada.
\newblock {\em Travel Behaviour and Society}, 16:161--170, 2019.

\bibitem{song2021visualizing}
Y.~Song, S.~Ren, J.~Wolfson, Y.~Zhang, R.~Brown, and Y.~Fan.
\newblock Visualizing, clustering, and characterizing activity-trip sequences via weighted sequence alignment and functional data analysis.
\newblock {\em Transportation Research Part C: Emerging Technologies}, 126:103007, 2021.

\bibitem{chen2019clustering}
R.~Chen, J.~Zhang, N.~Ravishanker, and K.~Konduri.
\newblock Clustering activity–travel behavior time series using topological data analysis.
\newblock {\em Journal of Big Data Analytics in Transportation}, 1(2):109--121, 2019.

\bibitem{stoffer1991walsh}
D.~S. Stoffer.
\newblock Walsh-fourier analysis and its statistical applications.
\newblock {\em Journal of the American Statistical Association}, 86(414):461--479, 1991.

\bibitem{harmuth1971transmission}
H.~Harmuth and J.~D. Lee.
\newblock Transmission of information by orthogonal functions.
\newblock {\em IEEE Transactions on Systems, Man, and Cybernetics}, (2):188--188, 1971.

\bibitem{shanks1969computation}
J.~L. Shanks.
\newblock Computation of the fast walsh-fourier transform.
\newblock {\em IEEE Transactions on Computers}, 100(5):457--459, 1969.

\bibitem{pereira2015persistent}
C.~M. Pereira and R.~F. de~Mello.
\newblock Persistent homology for time series and spatial data clustering.
\newblock {\em Expert Systems with Applications}, 42:6026--6038, 2015.

\bibitem{bubenik2015statistical}
P.~Bubenik.
\newblock Statistical topological data analysis using persistence landscapes.
\newblock {\em Journal of Machine Learning Research}, 16(1):77--102, 2015.

\bibitem{carlsson2009topology}
G.~Carlsson.
\newblock Topology and data.
\newblock {\em Bulletin of the American Mathematical Society}, 46(2):255--308, 2009.

\bibitem{zhang2021time}
Y.~Zhang, Q.~Shi, J.~Zhu, J.~Peng, and H.~Li.
\newblock Time series clustering with topological and geometric mixed distance.
\newblock {\em Mathematics}, 9(9):1046, 2021.

\bibitem{allahviranloo2017modeling}
M.~Allahviranloo, R.~Regue, and W.~Recker.
\newblock Modeling the activity profiles of a population.
\newblock {\em Transportmetrica B: Transport Dynamics}, 5(4):426--449, 2017.

\bibitem{corpet1988multiple}
F.~Corpet.
\newblock Multiple sequence alignment with hierarchical clustering.
\newblock {\em Nucleic Acids Research}, 16(22):10881--10890, 1988.

\bibitem{needleman1970general}
S.~B. Needleman and C.~D. Wunsch.
\newblock A general method applicable to the search for similarities in the amino acid sequence of two proteins.
\newblock {\em Journal of Molecular Biology}, 48(3):443--453, 1970.

\bibitem{zhu2009initializing}
Y.~Zhu, J.~Yu, and C.~Jia.
\newblock Initializing k-means clustering using affinity propagation.
\newblock In {\em 2009 Ninth International Conference on Hybrid Intelligent Systems}, 2009.

\bibitem{calinski1974dendrite}
T.~Caliński and J.~Harabasz.
\newblock A dendrite method for cluster analysis.
\newblock {\em Communications in Statistics-theory and Methods}, 3(1):1--27, 1974.

\bibitem{yin2021exploring}
B.~Yin and F.~Leurent.
\newblock Exploring individual activity-travel patterns based on geolocation data from mobile phones.
\newblock {\em Transportation Research Record}, 2675(12):771--783, 2021.

\end{thebibliography}

\end{document}